# Complex Networks and Waveforms from Acoustic Emissions in Laboratory Earthquakes

H.O. GHAFFARI, [1(a)] B.D. THOMPSON[2] and R.P. YOUNG [1]

(a)E-mail: _h.o.ghaffari@gmail.com_

[1] _Department of Civil Engineering and Lassonde Institute, University of Toronto, Canada, 170 College Street ,M5S 3E3,ON, Canada_
[2] _Mine Design Engineering, Kingston, Canada_

**Abstract** —We report some new applications of functional complex networks on acoustic emission waveforms from frictional interfaces. Our results show that laboratory faults undergo a sequence of generic phases as well as strengthening, weakening or fast-slip and slow-slip leading to healing. Also, using functional networks, we extend the dissipated energy due to acoustic emission signals in terms of short-term and long-term features of events. We show that the transition from regular to slow ruptures can have an additional production from the critical rupture class similar to the direct observations of this phenomenon in the transparent samples. Furthermore, we demonstrate detailed sub-micron evolution of the interface due to the short-term evolution of rupture tip, which is represented by phenomenological description of the modularity rates. In addition, we found nucleation phase of each single event for most amplified events follows a nearly constant time scale (about 4-8μs), corresponding to initial strengthening of interfaces.

**1) Introduction -** The 2011 Tohoku-Oki earthquake ($M_w$=9.0) showed how a slow slip phenomenon can lead to destructive ruptures with emerging successive rupture transitions from slow to sub-Rayleigh and super-shear ruptures [1-3]. Such a dramatic transition from creep fronts or slow slip events (SSE) to fast and sometimes critical ruptures has been reported in laboratory earthquakes: slow phase to fast, slow to critical phase and regular (fast) ruptures to super shears have been examined in transparent and rock materials [4-8]. The theoretical studies of rupture transition are focused mostly on regular rupture to super shear. The main controlling parameters of the models are characteristic length of the initial rupture and critical value (i.e., seismic ratio) regarding friction force [9-11]. Recently, a few numerical models have shown the evolution of creeping faults or SSE into critical earthquakes [12-13]. These models use the coupling of friction laws with heat and pore pressure. The assumption of releasing massive thermal energy is a vital part of the employed coupled equations of the models. In these models, the weakening of the frictional–rock interface (i.e., fault) is obviously due to rapid shear heating of pore fluids and does not accord with the aforementioned laboratory friction experiments. **Some other simpler models [14]-based on state-and-rate equations- capture essen-** **tials of rupture transitions in dry interfaces. However, these models are unable to predict all features of the laboratory observations. Thus, in the latter models, the real mechanism of velocity weakening or velocity strengthening in terms of friction evolution in micro and sub-micron scales are not known[6].**

In addition to these developments, recent precise laboratory measurements have presented new insights into rupture events with micro-seconds resolution [6, 15]. **Among these findings, universal trends of appropriate parameter spaces (such as displacement-time or temporal contact areas) were of considerable interest [6]. While short-term evolution of ruptures (and then friction) can be fairly demonstrated by universality of temporal strain (or velocity) profiles, the long term evolution of an interface is characterized by distribution of appropriate parameters. Among the long-term attributes, one can consider distribution of waveforms' amplitudes (i.e., b-value), distribution of waiting times between successive events and long-term evolution of healing parameters (such as contact areas).**

To support the aforementioned ideas, the authors used complex network techniques to extract "hidden" information from recorded waveforms as well as reconstructed or recorded images through simple friction-tests of glassy and rock samples [7]. We showed that functional networks constructed over sub-micron events ("precursor or foreshocks" laboratory earthquakes) can unravel the possible regime of the rupture while the changes of dynamic phases of network parameters are correlated with weak to very powerful events(i.e., energy spectrum of events) [7]. Here, using the proposed functional friction networks, we explore the transition regimes of ruptures in Westerly granite-frictional interfaces. To this aim, we use several network parameter spaces while the energy of the obtained networks (and then acoustic pulses) is analyzed using modularity profiles. Interestingly, we find that our phenomenological formulation based on the functional friction networks presents a solution to dramatic rupture transitions in laboratory scale. Building on the evolutionary phases of modularity's indexes, we also present a new way of estimating fracture energy where a glass transition period is matched with a rapid evolutionary phase of module fast-growth. Our results from the trends of fracture energy in different regimes of ruptures confirmed the recent theoretical and numerical studies regarding the physics of fracture energy. Furthermore, a slip-weakening model, constructed upon the frictional resistivity of the interfaces, is proposed where the resistivity-slip rate phase space is related to our networks' attributes. To complete our analysis, we developed a simple mean-field approach on the power law distribution of centrality, which lead to a phase diagram with a fit-get-rich-phase while we change the control parameter of the model. We show how the model can reach a gel-like or condensed state and interpret the "gel-like" functional friction networks as the "super-lubricity" or super-ruptures regime.

**2) Reviews of experiments -** The detailed experimental results have been reported in [16-17]. Our main data set includes the recorded discrete and continuous waveforms (i.e., acoustic emissions-AEs) using 16 piezoelectric transducers from a saw-cut sample of Westerly granite (LabEQ1), under triaxial loading [16-17]. The saw cut was at a 60 degree angle and polished with silicon carbide 220 grit. Each triggered event had the 48  duration of 204.8 µs (recorded at 5 MHz) while the three main stick-slip events occurred. The experiment was

49  servo-controlled using an axial strain rate of $5 \times 10^{-6} s^{-1}$ ($\sim 10 \mu m/s$ as the loading rate). The confining stress

50  was maintained at 150 MPa for three reported main stick-slip events, producing 109 located- rupture fronts.

51  The second data set (LabEQ2) are the results of the two main cycles of loading–unloading (stick-slip) of

52  Westerly granite on a preexisting natural fault by loading at constant confining pressure. A natural rough fault

53  was created using a triaxial loading system with constant confining pressure of 50 MPa and acoustic emission

54  feedback control.

55

56  **3) Functional-friction networks** – **The application of recurrence networks as new tools to analyze**

57  **time-series has been the subject of the numerous research areas during last decay [18]. In particular**

58  **building networks over recorded time series from different recorded instruments (receivers or observ-**

59  **ers) were of interest in analysis of time series from brain activity [19], climate networks [20], Confor-**

60  **mational dynamics of proteins [21] and network physiology [22]. Many methods for analysing interac-**

61  **tions between two or more time series have been proposed [23-26]. A recent proposed recurrence net-**

62  **work's meta-time series considers the power of the multi-observation of an event or series of events**

63  **through different observatories [27]. We recently employed a similar meta-time series analysis on the**

64  **quasi-static evolution of apertures and real-time contact areas [28-29]. Here, we describe a network on**

65  **each time step in which the nodes correspond with acoustic sensors where any elastic excitement induc-**

66  **es voltage-fluctuations in them. We show that the nature of piezoelectric signals can be quantified**

67  **through this new tool, resembling main features of instabilities in sub-micron scales. It notes that the**

68  **"functional" friction networks are not structural networks; however underlying functionality of con-**

69  **structed networks correlates with main mechanical features of friction instabilities. We use the word of**

70  **"functional" to distinguish our networks from the possible ones constructed over particles interactions**

71  **(such as force networks in granular materials).**

72  **Here we review the previously introduced algorithm. The algorithm includes the following five steps:**

73  **(1) Normalization of waveforms in each station.**

74  **(2) Each time series is divided according to maximum segmentation; we consider each recorded point in each**

75  **waveform with the length of $t_{max.} = 204 \mu s$. The $j$th segment from $i$th time series ($1 \leq i \leq N$) is denoted by $x^{i,j}(t)$.**

76  **We put the length of each segment as unit. This essentiality considers the high temporal- resolution of the system's**

77  **evolution, smoothing the raw signals with 20-60 times window (for Lab.EQ1& 2 is equal to 1-3 $\mu s$).**

78  **(3) $x^{i,j}(t)$ is compared with $x^{k,j}(t)$ to create an edge among the nodes. If $d(x^{i,j}(t), x^{k,j}(t)) \leq \xi$, we set $a_{ik}(j)=1$ oth-**

79  **erwise $a_{ik}(j)=0$ in which $a_{ik}(j)$ is the component of the connectivity matrix and $d = \left\| x^{i,j}(t) - x^{k,j}(t) \right\|$ is the**

80  **employed similarity metric.**

81     **(4) Threshold level ($\xi$): To select a threshold level, we use betweenness centrality (B.C). The details of the meth-**
82     **od have been explained in [28-29], and it has been proven that using this method quantitatively is equal to using**
83     **edge density.**

84     **(5) Increase the resolution of visualization: To decrease the sensitivity of the networks and knowing that recur-**
85     **rence networks generally reveal a good performance in a small number of nodes [19-20], we increased the size of**
86     **the adjacency matrix with the simple interpolation of $d$ using cubic spline interpolation. Further analysis shows**
87     **that the method is nearly insensitive for N>15. The increase in the number of nodes to 50 generally did not change**
88     **the presented results for the acoustic-friction networks and merely increased the visual quality of the results.**

89        Through this research, we use some of the main key networks' measures. We will use some of these
90     metrics to introduce phase-diagrams and further interpretation of obtained results. Each node is characterized
91     by its degree $k_i$ representing the number of links, and the betweenness centrality (B.C) [30]:

$$B.C_i = \frac{1}{(N-1)(N-2)} \sum_{\substack{h,j \\ h \neq j, h \neq i, j \neq i}}^{N} \frac{\rho_{hj}^{(i)}}{\rho_{hj}}, \qquad (1)$$

93     in which $\rho_{hj}$ is the number of the shortest path between $h$ and $j$ , and $\rho_{hj}^{(i)}$ is the number of the shortest path
94     between $h$ and $j$ that passes through $i$. For scale free networks whose degree distribution follows a power law
95     ,the distribution of B.C (or load) of a node-also- follows a power law [54]. In the last section, we will use the
96     distribution of centrality for each event (in ~200μs time interval) to form the component of a mean field mod-
97     el. **To describe the correlation of a node with the degree of neighboring nodes, assortatitive mixing in-**
98     **dex is used:** $r_k = \dfrac{< j_l k_l > - < k_l >^2}{< k_l^2 > - < k_l >^2}$ **, where it shows the Pearson correlation coefficient between de-**
99     **grees** $(j_l, k_l)$ **and** $< \bullet >$ **denotes average over the number of links in the network [30].**

100     **The network's modularity characteristic is addressed as the quantity of densely connected nodes rela-**
101     **tive to a null model (random model). Based on the role of a node in the network modules or communi-**
102     **ties, each node is assigned to its within-module degree ($Z$- scores) and its participation coefficient (P).**
103     **High values of $Z$ indicate how well-connected a node is to other nodes in the same module, and P is a**
104     **measure of the well-distribution of the node's links among different modules [32].**

105     **The main celebrated diagnostic in our networks is $Q$-profiles.** The modularity $Q$ (i.e., objective function
106     which we are looking to maximize) is defined as [32]:

$$Q = \sum_{s=1}^{N_M} [\frac{l_s}{L} - \left(\frac{d_s}{2L}\right)^2], \qquad (2)$$

108    in which $N_M$ is the number of modules (clusters), $L = \frac{1}{2}\sum_i^N k_i$, $l_s$ is the number of links in modules and

109    $d_s = \sum_i k_i^s$ (the sum of node degrees in modules). We use the Louvian algorithm to optimize Eq.2 [31], which

110    has been used widely to detect communities in different complex networks. **Then, in each time step during**

111    **evolution of waveforms (here ~200 μs), we obtain a Q value . The temporal evolution of Q values in the**

112    **monitored time interval forms Q-profile. Q-profiles carry generic universal time scales, likely distingush-**

113    **ing microseconds details of micro-cracks. Regarding wide range applications of acoustic emissions and**

114    **their observation in  an unusually large number of  experiments (such as fractures ,martensite trans-**

115    **formation, lattice dislocations, high-pressure/high-temperature experiments leading to phase transfor-**

116    **mation of minerals , etc), finding such a universal frame on complicated emitted waveforms are much of**

117    **important. This likely presents a road to draw a universal picture of acoustic emission sources, close to**

118    **mechanisms of breaking or deforming of  atomic bonds.**

119    Furthermore, we will use clustering coefficient as the index of normalized triangles in each existing node.

120    The number of triangles surrounding a node in the constructed networks, denoted by $T_i$, is used to define the

121    clustering coefficient:  $c_i = \frac{2T_i}{k_i(k_i - 1)}$.   Furthermore, the network's spectrums or several parameter spaces

122    based on the introduced measures have been widely employed to analysis the structures and the possible phys-

123    ical mechanisms behind the obtained networks. **Parameter spaces** such as *c-k*, *C-B.C* (or *k*-B.C) have been

124    used widely in analysing a wide range of networks [26, 29, 32-33].

125

126    **4) Network's Parameter Spaces of AEs Events –** The algorithm results a typical Q-profile for each sin-

127    gle event (Fig.1a). This profile presents some basic ingredients of a single event as the result of breaking an

128    asperity. In the following, we show how temporal evolution of the  modularity index represents a generic dy-

129    namic of a single asperity failure, including information on nucleation, fast-deformation and fast-slip parts.

130    Inspecting over 8000 events from Lab.EQ1 and Lab.EQ2 revealed that the general evolution of Q-profiles is

131    universal, imprinting nearly constant characteristics of timescale per each evolutionary phase [7,15]. The main

132    evolutionary phases are as follows: phase I is the rapid dropping of modularity and reach to the minimum val-

133    ue of Q(t), indicating the formation of  a dense  network (i.e., jamming phase). Considering that the laboratory

134    earthquakes are in near-field, approaching the crack front, deformation or crushing the "node" or asperity is

135    the possible explanation for this stage. The duration of this phase is 4-8  $\mu s$ . **We note that this phase corre-**

136    **sponds with an initial strengthening of fault, which impose an obstacle for rupture growth**. **The dura-**

137    **tion of this phase is much smaller than the other following generic phases. Sudden release of energy -**

138 **stored in this a few micro seconds- likely dictates rupture dimension and rupture regime.** Later, we will

139 use the inverse of the Q-profiles to magnify the first phase as the nucleation-deformation phase.

140 The second phase is the fast growth of modularity and increase in the number of modules as has been

141 shown in Fig.1.a. As we have shown in [7], the rate of this rebounding phase is scaled with the maximum val-

142 ue of the modularity. The collapsed data set for this phase indicated a universal generic dynamic, which is

143 confirmed by a constant value of the power law coefficient $\alpha_2 \approx 1.5 \pm .3$ for both Lab.EQ1 and Lab.EQ2. The

144 duration of this phase is 25-35 $\mu s$. **The possible approach for this phase is the fast slip of the interface**

145 **which is arrested or damped at the end of the phase. The fast slip stage or fast-weakening phase coin-**

146 **cides with unjamming of modules. It is important to understand why the duration of this phase is near-**

147 **ly constant for both experiments while the general loading conditions were different. We will show this**

148 **time scale is different in cement (concrete). Then, we infer that fast-weakening phase is a characteristic**

149 **of tiny ruptures with strong dependency on molecular /atomic structure of materials.** Assuming a range

150 of velocity for phase II, 5-500 $\mu m / s$, we obtain the maximum displacement ~150pm-15nm. This simple cal-

151 culation shows that tiny-amplified events are closely related to failure of atomic bonds of heterogeneities and

152 asperities. From this perspective, emitted noises from failure of chemical bonds in vicinity of a crack tip

153 should be strongly scaled with the molecular structure and forces between them (leading to cohesive forces or

154 surface energy). The maximum displacement at the end of the phase 2 changes to 1-5 $\mu m$ if the frictional slip

155 rate increases to 50-100 $mm / s$ (the reported value for PMMA in [15]). This shows that the recorded precur-

156 sor events encode a displacement magnitude at the range of pico to nano meter (and maximum micro),

157 demonstrating possible nano-earthquakes in terms of displacement of the interface.

158 The transition to the last evolutionary phase (phase IV) is accompanied by an unusual gradual decay

159 of the communities (phase III). We will use a theory of glass transition with employing an effective tempera-

160 ture model to explain the emergence of this phase (see next section). The last stage is generally a decaying

161 phase, indicating a slow-slip phase. The rate of the slow slip phase, $\dot{Q}_{IV}$, scales with the maximum recorded

162 modularity at the following rate: $\left| \dot{Q}_{IV} \right| \propto Q_{max}^\lambda$ in which λ is nearly constant for the regular events (supplemen-

163 tary document of [7]). **Regular events are events with rupture speed under a threshold level (Rayleigh**

164 **velocity) .** Furthermore, a rough scaling between the rate of the first phase and the last phase was found [7]:

165 $\left| \dot{Q}_{IV} \right| \propto \left| \dot{Q}_I \right|^{-\gamma}$ where γ is slightly larger for slow slip events. To get a spatio-temporal evolutionary picture of

166 the recorded events (**and then relatively long-term evolutionary patterns of ruptures**), we use mean values

167 of network diagnostics. Applying this idea to Lab.EQ1 and Lab.EQ2, we recognized several correlations

168 among network attributes (Fig.1b,c *and* Fig.2a).

169

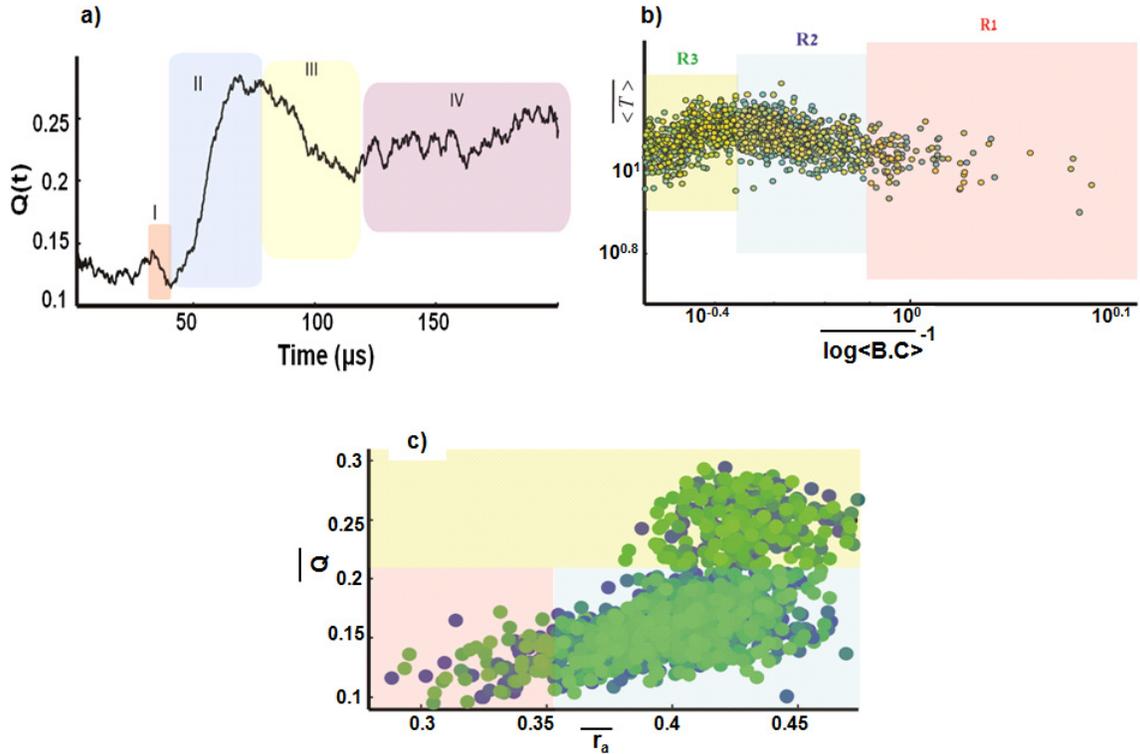

**Figure 1| Evolution of rupture fronts in a rough fault.** (a) Q-profile resulted from the mapping of acoustic waveforms form a regular event includes 4 main evolutionary phases [7]. Two sharp transitions can be followed from phase I to II and from II to phase III. (b) The scaling of $\overline{\log <B.C>}^{-1}$ with the temporal average of triangles (3 points loops: $\overline{<T>}$) differentiates the three main clusters: $\overline{<T>} \propto \overline{\log <B.C>}^{-x}$. Also see Fig.2.a; (c) Higher average assortativity scales with higher modularity value, with a separation in high modular fronts, characterizing slow slip ruptures. The colors in panels b and c show the events' sequences, corresponding to time.

The parameter spaces shown in Figures 1b, c and Fig.2.a demonstrate three main classes. In Fig.2a, we show one of the parameter spaces for Lab.Eq.2 : $\overline{\log <B.C>}^{-1} - \overline{Q}$. For relatively high values of $\overline{\log <B.C>}$, a crossover from high-modularity to intermediate modularity is observed. For the saw-cut experiment, we found a unique separation of some events, providing a slow deformation (i.e.,R3 class) with respect to the duration of phase I and somehow a similarity to the phase IV [7]. A similar crossover is observed in $\overline{<T>} - \overline{\log <B.C>}^{-1}$ and $\overline{Q} - \overline{r_a}$ (Fig.1b, c), in which $\overline{<T>}$ is the spatio-temporal mean of triangles . We recognize three main classes regarding this network-clustering: R1, R2 and R3. To classification of phase spaces, the trends of clusters are considered. **Obviously, one can also use simple clustering methods such as K-means clustering, Self-organizing feature map or other data clustering algorithms.** The R1 phase holds events with relatively high energy, while events with the intermediate energy ("regular") are allocated to the R2 class. The R3 class does have a longer evolutionary phase I and represents longer and weaker (on average)

189 events, indicating a slow-lip phenomenon. Based on Fig.1b and Fig.2a we can write a power law function to

190 cover the collapsed events: $\overline{<T>} \propto \overline{\log <B.C>}^{-\chi}$ and $Q_{max} \propto \Xi^{-\pi}$ in which $\Xi = \overline{\log <B.C>}^{-1}$, π and χ are

191 nearly constant exponents for each class. Obviously, transition from R3 to R2 class is described with change

192 of the sign of χ from positive to negative.

193

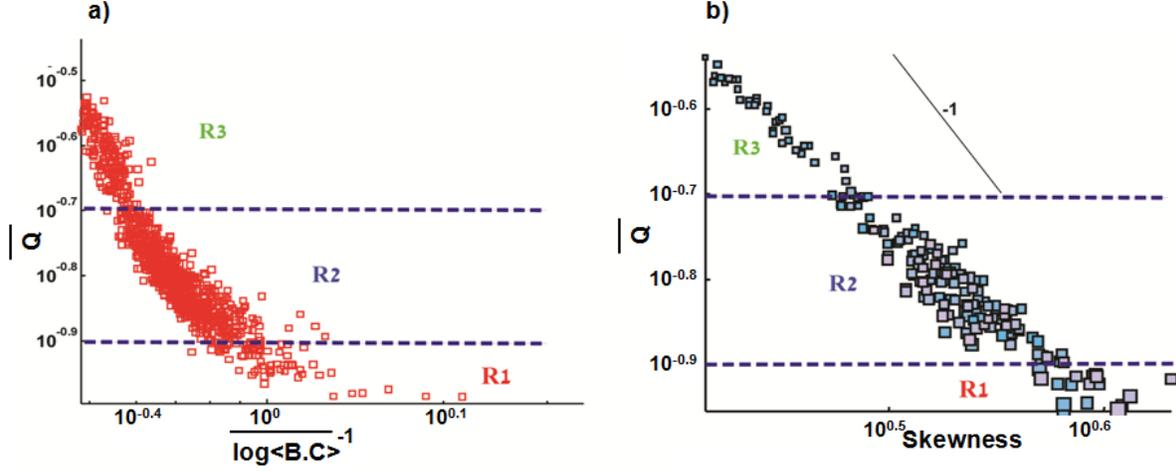

195 **Figure 2| Asymmetric nature of Q-profiles. (a)** $\overline{\log <B.C>}^{-1} - \overline{Q}$ **plane with three distinct classes. (b) The average skewness**

196 **scales with** $\overline{Q}$ **, indicating a smaller positive skewness of pulse shape holds events from R3 class (data from Lab.EQ2). The size**

197 **of squares corresponds with the intensity of** $\overline{\log <B.C>}^{-1}$ **and the colors show the sequence of events.**

198

199 Furthermore, we can investigate the asymmetric nature of crackling noise pulses in terms of Q-

200 profiles. The average skewness of Q-profiles as the asymmetry measure can be quantified **by [Eq.1 in 35]:**

$$\Sigma = \frac{\frac{1}{t_{max.}} \int_0^{t_{max}} Q(t)(t - \overline{t})^3 dt}{[\frac{1}{t_{max.}} \int_0^{t_{max}} Q(t)(t - \overline{t})^2 dt]^{3/2}}, \tag{3}$$

202 in **which** $\overline{t} = \frac{1}{t_{max.}} \int_0^{t_{max}} dt Q(t)t$ and $t_{max.} = 204 \mu s$ . In Fig.2.b, we have compared the skewness versus the

203 mean modularity for the Lab.EQ2 events. An excellent collapsing of data in $\overline{Q} - \Sigma$ shows an inverse correla-

204 tion of the leftward asymmetric shape of the acoustic waveforms regarding the regime of ruptures. This is a

205 universal feature of "crackling noise" systems which has been allocated to the nature of the dissipation energy

206 such as eddy currents in Barkhusen noise (movement of magnetic domain wall) or threshold strengthening in

207 moment rate profiles in natural earthquakes [34-35]. We conclude that more deviation from symmetry is the

208 signature of ruptures with relatively high energy and critical ruptures, while approaching a less asymmetric

209     shape indicates ruptures with lower energy. **Then, understanding the details of micro-seconds evolution of**

210     **Q-profiles helps to evaluate the general shape of crackling noises.**

211        Next, to distinguish the role of main deformation phase I, we define a parameter of resistivity against

212     motion, denoted by resistivity: $R \equiv \dfrac{1}{Q_{norm.}}$ where $Q_{norm.} = \dfrac{Q}{Q_0}$ (**$Q_0$ is the rest value of Q(t)**) In Fig.3a, we

213     have shown R-profiles from Lab.EQ1. In Fig.3c, we have also illustrated the collapsing of events from

214     Lab.EqQ1 in a normalized space : $(\dot{R})_I - R_{max.}$, indicating a nearly constant time scale for the first phase and

215     for *most* of the events. We calculated this constant time as ~2-5μs, indicating that nucleation phase leading to

216     the main failure of the asperity is a nearly constant and universal (for regular events) generic dynamic. Ap-

217     proaching to the critical stage and subsequent fast-slip phase can be well-quantified by "$S_R$-parameter":

218     $S_R = \dfrac{R_{max.} - R_0}{R_0 - R_{min.}}$ (Fig.3a-inset). This parameter is similar to seismic S-factor: $S_s = \dfrac{\tau_p - \tau_0}{\tau_o - \tau_r}$ **[9-11] which is**

219     **defined over shear strength variation in a period of shear displacement, where** $\tau_p$**,** $\tau_0$ **and** $\tau_r$ **are max-**

220     **imum, initial and steady state shear strength, respectively.** It has been widely discussed that the value of

221     $S_s$ could be used to determine the regime of rupture; for instance it has been shown that a super-shear rupture

222     may occur below a critical value of $S_s < S_c$. In other words, the probability of occurring "critical ruptures"

223     decreases, and the regime of the rupture approaches regular and slow ruptures by increasing $S_s$. In Fig.3d, we

224     have shown that $S_R$ scales with $Q_{max.}$ representing ruptures with smaller maximum modularity hold higher

225     $S_R$. On the other hand, generally smaller $Q_{max.}$ indicates weaker and slower failures in sub-micron scales.

226     Consequently, we can use the "$S_R$-parameter" in the possible classification of rupture fronts (Fig.3d).

227

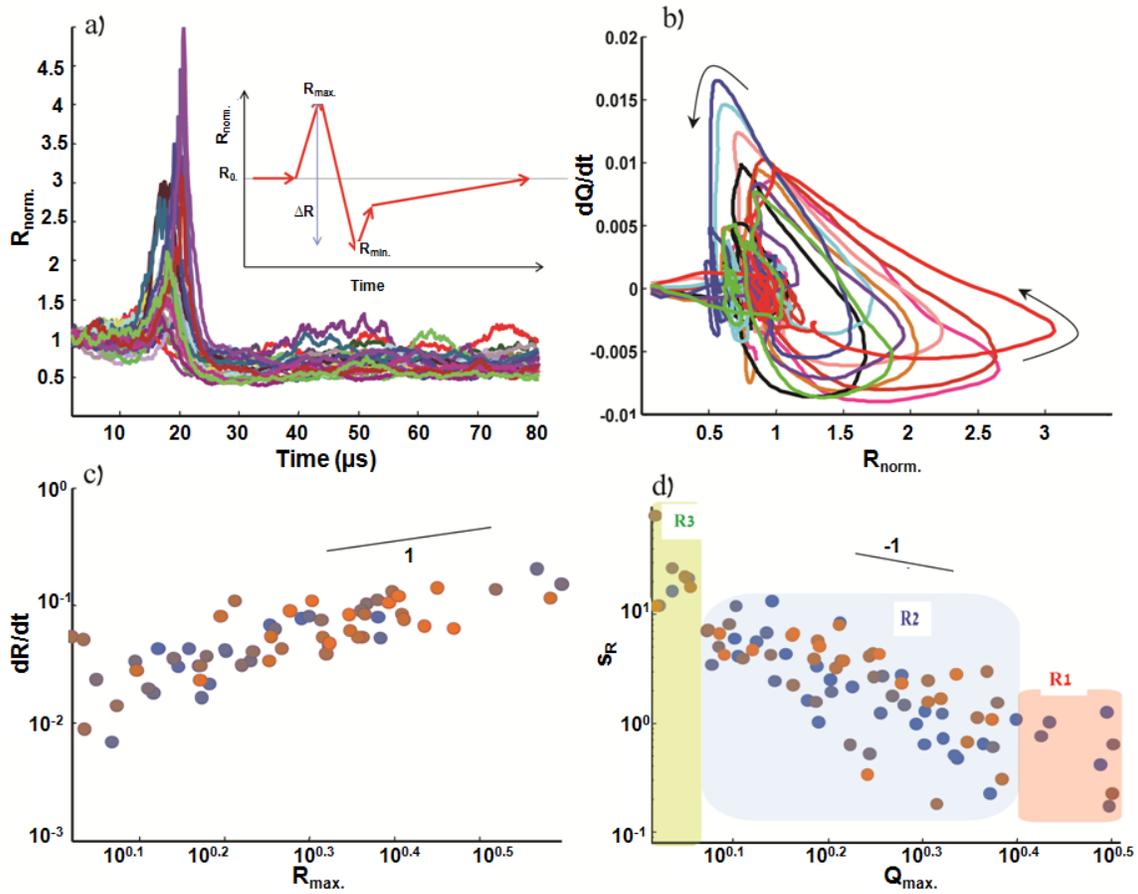

**Figure 3| "Resistivity" against interface motion and the onset of slip- weakening in micro-second resolution** (a) The inverse of $Q$-profiles amplifies the evolution of the first phase. We assume the resistivity against the slip (or motion) ranges from proportional to reciprocal of the Q values.   We have shown ~30 arrested rupture fronts from Lab.EQ1 and their normalized resistivity parameter (Inset: a simplified form of the resistivity profile parameters to define "S" parameter). (b) Map of events from Lab.EQ1 in $R_{norm.} - \dot{Q}$ plane (c) a nearly constant exponent ($_\mu$) in $\dot{R} \propto R_{max.}^{\mu}$ represents a generic dynamic for the first evolutionary phase, indicating a failure or deformation timescale. (d) Collapsing events in $S_R - Q_{max}^{norm}$ parameter space indicates events with higher modularity show smaller $S_R$ value, while very small Q values hold higher $\dot{S}_R$ values (events from Lab.EQ1). In panels (c) and (d), the colors show the sequence of events. Earlier events are blue and the final ones are orange. Slopes of 1 and -1 for the coefficient of a power law between the shown parameters are displayed for reference.

Next, we show that $R_{norm} - \dot{Q}$ plane reflects a similar acceleration –deceleration phenomenon in frictional interfaces under high velocity (Fig.3.b) [36]. Here $R_{norm}$ is the normalized value of resistivity profile. **To fit a model on $R_{norm} - \dot{Q}$ plane (Fig.3b and Fig.4), we use the fact that the duration of the second phase ("fast-slip" or slip-weakening) is nearly constant for "regular" ruptures. In addition to this, to fit a proper function we use studies in which  the slip rates of frictional interfaces are related into the fric-** 245 **tional resistivity and also involve the accelerating and decelerating of faults during fast-rate of slips**
246 **(such as Eqs.1 and 2 in [36]). This leads to:**

247
$$R = R_0 + A(R_p - R_0) \exp[\ln(0.03) \; t/t_c], \tag{4}$$

248 **in which** $R_0(\dot{Q}) = \exp(-\dot{Q}/\dot{Q}_c)$ **and** $R_p$ **is the maximum of resistivity (***t*** is time). We found that**

249 $t_c \approx 70 \mu s, \dot{Q}_c = 0.004, A = 4$ **satisfies most of the recorded experimental results (Fig.4). Essentially the**

250 **presented relation involves two main controlling parameters:** $R_p$ **and** $t$. **Considering that** $t_c$ **includes**

251 **some dead time and the first phase duration in Q-profiles, we estimate** $(t_c)_{real} \approx t_{II}$ **,** *i.e.,* **weakening time**

252 **or fast slip duration. As if Eq.4 is in time domain (and not in space such as displacement); however,**

253 **Eq.(4) presents a quite similar phase space to** $\mu_{friction} - \dot{u}$ **(friction-slip rate) in the mentioned reference.**

254 **We rationalize this similarity to micron (and sub-micron rupture fronts) with considering time-**

255 **weakening constitutive law, which prescribes that the fault's resistance weakens with increasing time**

256 **over a characteristic time (see [37] for more information).**

257

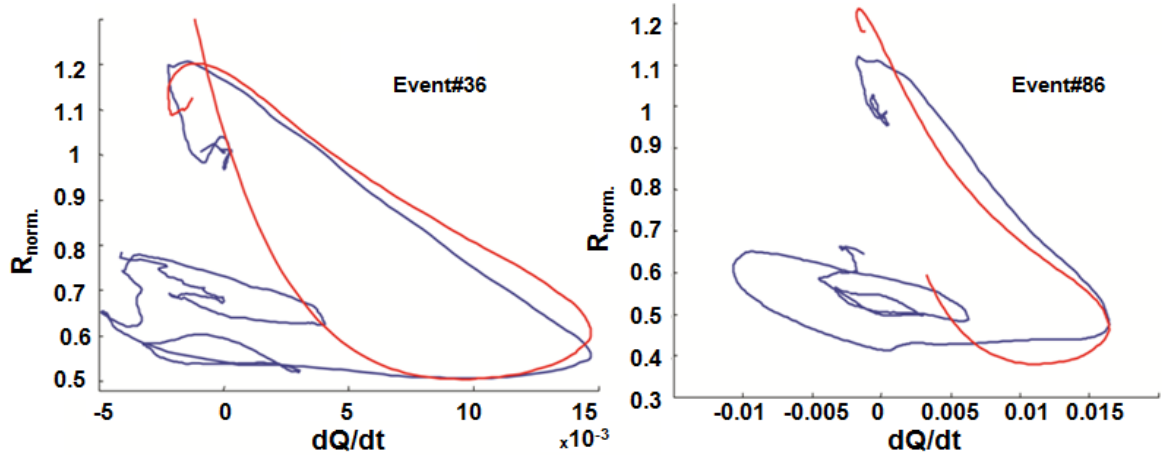

259 **Figure 4** | Comparing Eq.4 with the experimental data from Lab.EQ1. A comparison of phase space of precursor rupture fronts
260 from smooth-rock interfaces in terms of rate of Q- profiles and resistivity between experimental data (blue) and friction modeled by
261 presented equation (red-dashed line).

262

263

264

265

266

267

**5) Rupture-Energy Based on Modularity Rate –** In this section, we will use some of the mentioned relations to establish a possible link with energy of ruptures. Moreover, we estimate fracture energy based on the revealed evolutionary phases durations. **In fact, this assumption implies that acoustic energy release during a rupture phenomena may be assumed to be proportional to the dissipated energy. We develop an energy term for the fixed monitored time interval based on the rate of evolution of modules. We assume a friction network and its energy are evolved during a time interval and through the aforementioned distinct phases. We investigate the accumulation of energy of the system by calculating the rate of variation of communities in each generic phase. Then, we couple short-term parameters of an event (i.e., Q-profile) with long-term trends of distinct events. Also, we speculate that the energy of functional networks in terms of evolution of communities can be purely expressed in terms of 3-point closed loops (triangles). In [40-42], an analysis of motifs of different networks proved that the functionality of networks was correlated with the motif's profiles. If the energy of a system could be extended in terms of motifs, we can expect that the motif's frequency is associated with the energy spectrum of the system. Then, as if we did not extend our calculations on other types of motifs, however we conjecture that the building proper relations between motifs and other network parameters shall provide similar (and not identical) expression on energy of the system. It is noticed that the energy of networks has been studied previously in which an energy term was extended based on vertex degrees, global properties (such as sizes of network's component) and number of modules [33,38].**

286
287
288
289
290
291
292
293
294
295
296
297
298
299

303 **Table 1**. **List of parameters used in this study.**

304

| Parameter | Description |
|---|---|
| $k_i$ | Degree of a node or number of links for a given node. |
| $C$ | Clustering coefficient |
| $\Sigma$ | Skewness of signals |
| $\overline{<T>}$ | Spatio-temporal average of triangles .Spatial average is on all nodes and temporal mean is on the short-term monitored interval. |
| $\overline{<B.C>}$ | Spatio-temporal average of centrality |
| $Q_{max}$ | Maximum modularity value in the given time interval in which evolution of waveforms are monitored. |
| $\Xi$ | $\Xi \equiv \overline{\log <B.C>}^{-1}$ |
| $\dot{Q}_i$ | The rate of Q(t) per each generic phase in Q-profiles. |
| $R$ | R-profiles as the reciprocal of Q-profiles. We use this value to emphasis on the first evolutionary phase as the nucleation and main deformation phase of micro-cracks. |
| $\chi$ | Power exponent in $\overline{<T>} \propto \overline{\log <B.C>}^{-\chi}$. |
| $\alpha_2$ | Power exponent in $\dot{Q}_{II} \propto Q_{max}^{\alpha_2}$. |
| $\pi$ | Power exponent in $Q_{max} \propto \Xi^{-\pi}$. |
| $\lambda$ | Power exponent in $\left\|\dot{Q}_{IV}\right\| \propto Q_{max}^{\lambda}$. |
| $\gamma$ | Power exponent in $\left\|\dot{Q}_{IV}\right\| \propto \left\|\dot{Q}_I\right\|^{-\gamma}$. |
| $\mu$ | Power exponent in $\dot{R} \propto R_{max.}^{\mu}$. |
| $E$ | Sum of energy of the generic phases in Q-profiles. |
| $\kappa$ | Coefficient in $\overline{Q} \propto \exp(-\kappa F_{axial})$. |
| $\eta$ | Power exponent in $\overline{Q} \propto \Xi^{-\eta}$. |
| $F_{axial}$ | External axial load as the driving force of the system. |
| $\Gamma$ | Fracture surface energy; onset of cracking is satisfied when fracture energy meets Griffiths crack criterion. |
| $\Omega$ | Power exponent in $P(BC) \sim BC^{-\Omega}$. |
| $\xi$ | Power exponent in $<B.C>_x \propto <C>_x^{-\xi}$; $x$ stands for spatial nodes. |
| $\beta$ | Power exponent in $T(k) \propto k^{\beta}$. |
| $\delta$ | Power exponent in a scale free network $P(k) \propto k^{-\delta}$. |

310

311     **In table 1, we summarized the employed parameters in our calculations.** Considering

312 $\overline{<T>} \propto \overline{\log <B.C>}^{-\chi}$, $\left|\dot{Q}_{IV}\right| \propto Q_{max}^{\lambda}$ and $Q_{max} \propto \Xi^{-\pi}$ in which $\Xi = \overline{\log <B.C>}^{-1}$, we reach to : $\overline{<T>} \propto \left|\dot{Q}_{IV}\right|^{\frac{-\chi}{\pi\lambda}}$.

313 We approximate the kinematic energy of the system as the energy of each evolutionary phase (shown in

314 Fig.1a) in Q-profiles: $E_{system} \approx E_I + E_{II} + E_{III} + E_{IV}$. We also assume $E_{IV} \propto \dot{Q}_{IV}^2$, which leads to the following:

315 $E_{IV} \propto \overline{<T>}^{\frac{-2\pi\lambda}{\chi}}$. This assumption is due to the similarity of Q-profiles to the recorded strains (normalized

316 displacements) [6]. With ignoring the third phase, we estimate $E_I \propto \dot{Q}_I^2$ and we get the first and the fourth

317 terms in the energy term by using $\left|\dot{Q}_{IV}\right| \propto \left|\dot{Q}_I\right|^{-\gamma}$, leading to $E_{system} \propto \overline{<T>}^{\frac{-2\pi\lambda}{\chi}}(1 + \overline{<T>}^{\frac{(2\pi\lambda)(1+\gamma)}{\chi}} + E_{II})$. The last term

318 (the second term in $E_{system}$) is extended as follows: $E_{II} \propto \dot{Q}_{II}^2 \propto \overline{<T>}^{\frac{-1}{2\chi\pi\alpha_2}}$ where $\alpha_2$ is constant for *most* of the

319 events. Eventually, we reach the following:

320
$$E_{system} \propto \overline{<T>}^{\frac{-2\pi\lambda}{\chi}}\left(1 + \overline{<T>}^{\frac{(2\pi\lambda)(1+\gamma)}{\chi}}\right) + \overline{<T>}^{\frac{-1}{2\chi\pi\alpha_2}}, \tag{5}$$

321 in which $\{\pi, \lambda, \gamma, \alpha_2 > 0\}$, and it presents the kinematic energy of the pulse regarding Q-profiles. In other

322 words, the energy of the waveforms is extended with motif frequency and some exponents.

323

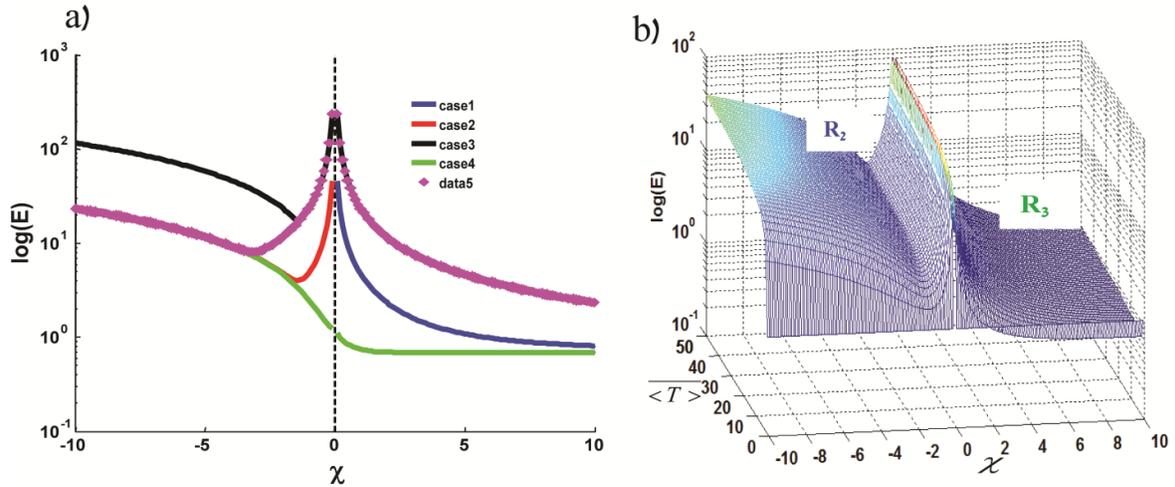

324

325 **Figure 5| Study of rupture transition using energy of Q-profiles.** (a) Variation of the energy with the exponent in $\overline{<T> \propto \log <BC>}^{-\chi}$

326 shows a singularity around $\chi = 0$ in which a transition from $R_3$ to $R_2$ occurs for $\overline{<T>}=10$. Case 1: $\pi = \lambda = \gamma = 1, \alpha = 2$; Case 2:

327 $\pi = \lambda = 1, \gamma = 5, \alpha = 2$; Case 3: $\pi = 2, \lambda = \gamma = 1, \alpha = 2$; Case 4: $\pi = \gamma = 1, \lambda = 0.01, \alpha = 2$; Case 5: $\pi = \gamma = 1, \lambda = 5, \alpha = 2$. (b) Energy varies with $\chi$

328 and $\overline{<T>}$ for the first case.

329

330       When $\chi > 0$ ($R_3$-regime-Fig.1b), increasing the loops pushes the system to a steady state with the min-

331     imum energy, while the increasing loops leads the system to a new energy level as the minimum upper bound

332     of $R_2$: $E_{system,\chi<0}^{<\overline{T}>\rightarrow \max.} \propto <\overline{T}>^{\frac{-2\pi\lambda}{\chi}} + <\overline{T}>^{\frac{-1}{2\chi\pi\alpha_2}}$. We denote this upper bound, as the crossover point from $R_3$ to

333     $R_2$ (see Fig.5.a). In other words, we have shown the kinematic energy in the crossover point ($R_3$ to $R_2$ and vice

334     versa) changes dramatically, reassembling the first order transition. The obtained result is similar to the direct-

335     observation of transition from sub-Rayleigh to slow rupture mode [4-8]. Approaching $\chi = 0$ does have an-

336     other interpretation, comparable with reaching to $R_1$ (or critical ruptures) class. Jumping or crossing from the

337     energy barrier as the singularity point of this phenomenological model can be interpreted as the generation of

338     a high energy rupture, i.e., a signature of $R_1$ class. This interpretation is similar with the observation of super-

339     shear rupture in PMMA – (reported in [4]; see also Fig.6). In [4] in fact, the transition from sub-Rayleigh rup-

340     tures to slow ruptures has been shown for the first time experimentally to result in a third production rupture

341     from super-shear rupture (Fig.6).

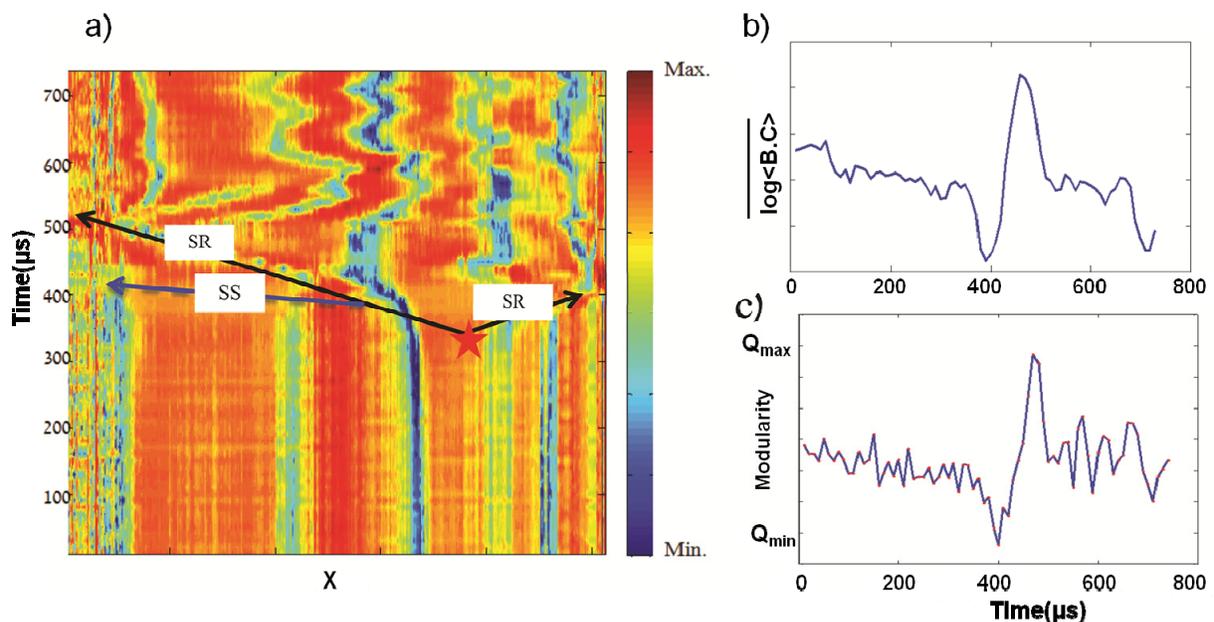

342

**Figure 6| Transitions in Laboratory Ruptures.** **(a)** Spatio-temporal variation of Z-score -as defined in [32]- over implementation of
networks on real-time photo elastic of successive photos from Resin-Resin interface [8]. We have used a method introduced in [28-29]
to construct networks over raw images. The transition to sub-Rayleigh (SR) ruptures does have a production from high energy rup-
tures, super-shear ruptures (SS). This result has been previously confirmed by Fineberg's group [4] .X=14.5 cm is the length of the
interface. **(b,c)** Mean centrality and $Q$ values for the shown rupture versus time in ~700µs time-window. The data set courtesy of the
original photo-elastic movie is from S.Nielson (INGV, Italy).

349

350       Assuming independence of scaling exponents to each other and approaching $\lambda \rightarrow 0$, $R_3$ class is van-

351     ished (Fig.5a-green line). We conclude that $\lambda, \pi, \gamma$ controls transitions from $R_2$ to $R_1$ or $R_3$, the sharpness of transition from R$_2$ to other classes and the evolution of R$_3$ class respectively. Remarkably, the transition from R$_2$ to R$_1$(or R$_3$) with approaching $\chi$ to $\chi = 0$ shows crossing from a minimum $E$, indicating a jammed regime of the rupture. The jammed state of the introduced-rupture's parameter space is important to understand because it can arrest the rupture as it propagates in the energy landscape. The same behavior is observed when $\gamma \to 0$ while the transition from R$_3$ to R$_1$ is trapped in a jammed state (not shown). Moreover, using the results presented in [7], we connect $\overline{<T>}$ to the remote stress (strain) field. To this end, we use $\overline{Q} \propto \exp(-\kappa F_{axial})$ and $\overline{Q} \propto \Xi^{-\eta}$, which yields $\log \overline{<T>} \propto \dfrac{\chi \kappa}{\eta} F_{axial}$ in which $\kappa > 0$ and $\eta$ are scaling exponents. $F_{axial}$ represents external axial loading stress as the driving force (boundary condition). For the R$_2$ class ($\chi < 0$), decreasing $\overline{<T>}$ corresponds with the increasing external remote stress field, while the trend of $\overline{<T>}$ is similar to $F_{axial}$ for $\chi > 0$. To get use of this scaling relation, we substitute $\log \overline{<T>} \propto \dfrac{\chi \kappa}{\eta} F_{axial}$ in to the energy term, yielding:

$$E_{system} \widetilde{\propto} (\frac{\chi \kappa}{\eta} F_a + 1)^{\frac{2\pi \lambda \gamma}{\chi}} + (\frac{\chi \kappa}{\eta} F_a + 1)^{\frac{-1}{2\chi \pi \alpha_2}} . \qquad (6)$$

In the next step, we summarize two regimes for the slow-fronts class (R$_3$): (1) when $\chi$ is high: $E_{system}^{R_3} \to const.$ and (2) when $\chi$ is small or $F_a$ is high: $E_{system}^{R_3} \propto (\frac{\chi \kappa}{\eta} F_a)^{\frac{2\pi \lambda \gamma}{\chi}}$. It is noteworthy that in the presented results, $\chi$ is treated as the order parameter.

As the last part of this section, we use an approach suggested for the left-hand asymmetric shape of the (average) of avalanches in crackling noise systems. Based on this approach [34-35], the asymmetric average shape of the avalanches is due to the role of energy dissipation phenomena (eddy currents in Barkhausen noise and strengthening threshold in seismic moment signals). Here, we use an equal version for the energy dissipation phenomenon in frictional interfaces, originally proposed in [7,15] to explain the abnormal drop of the phase III (Fig.1a). The interpretation is as follows: a fast-short time fracturing (phase I) induces a very fast increase in the temperature of a tiny "process zone", which cools in typical time characteristics. **The main component of the theory is that the whole of the fracture energy is transferred to heat in the process zone (Fig.8). This can be assumed as an effective temperature approach, well described in annealing-embrittlement phenomena [39]. Based on this theory, the annealing period is reflected in the form of fast slip phase, and the immediate strengthening stage is embrittlement (i.e., phase three in Q-profile). The increased temperature with respect to a reference temperature (room temperature) for 1D case with an imminent released energy as the source is given by [6]:**

$$\Delta T_{temp.} = \frac{-\Gamma}{4\rho c_p h}[erf(\frac{-h}{\sqrt{4D_T t}}) - erf(\frac{h}{\sqrt{4D_T t}})], \qquad (7)$$

in which $erf(x) = \frac{2}{\sqrt{\pi}}\sum_{n=0}^{\infty}\frac{x}{2n+1}\prod_{k=1}^{n}\frac{-x^2}{k} \approx \frac{2}{\sqrt{\pi}}(x - \frac{x^3}{3} + \frac{x^5}{10} - ...)$, $h$ is the thickness of the process zone in which the en-ergy rapidly dissipates, $D_T$ is the thermal diffusivity , $\Gamma$ is the energy released in phase I (or fracture energy),

and $t$ is the cooling time. With first-order approximation, we estimate $t$ as follows:

$$t \approx \frac{\Gamma^2}{\Theta}, \Theta = 4(\Delta T \rho c_\rho)^2 \pi D_T \qquad (8)$$

in which $\Theta$ is a constant value for a given material. From the previous section, we assume $t \approx t_{II}$. Since

$\dot{Q}_{II} \propto \frac{1}{t_{II}}$, we then discover: $\dot{Q}_{II} \propto \frac{\Theta}{\Gamma^2} \Rightarrow E_{sys.}^{II} \propto \Gamma^{-4}$. Then the $E_{pulse}$ indirectly includes the fracture energy.

The implication of the obtained relation is remarkable when we analyse the fracture energy variation while considering the rupture velocity-regimes: $E_{sys.}^{II} \propto \overline{<T>}^{-1/2 \chi \pi \alpha_2}$ yielding $\Gamma \propto \overline{<T>}^{1/8 \chi \pi \alpha_2}$. Considering $\chi < 0$ indicates that decreasing $\overline{<T>}$ results in the increment of the rupture velocity (i.e., $\overline{<T>} \propto \Xi^\chi$ parameter space), inducing increment of $\Gamma$.

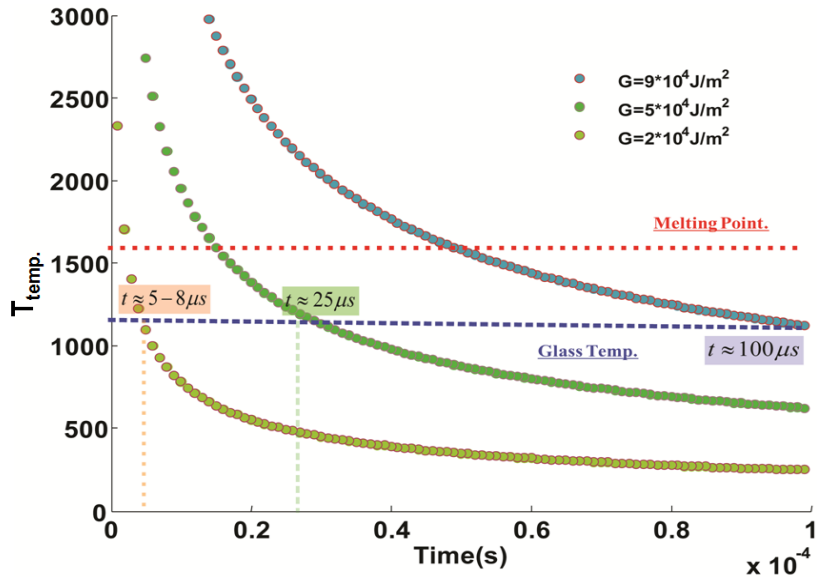

**Figure 7|Transition to "rubber-like" phase using 1D heat-diffusion model for a sub-micron process zone.** For $G \approx 5 \times 10^4 \ Jm^{-2}$
[47], the duration of the second phase $t \approx 25 \mu s$ is satisfied. **Temperature is in degree Celsius (°C).**

The obtained result is comparable with the results of several investigations regarding the weak in-crease of $\Gamma$ with rupture velocity [43-44]. We also can extend the concept of the rate of energy dissipation in 398    a "slow class" regime (R3). In other words, how does fracture energy change in slow-slip regime? In this case,

399    $\chi > 0$ and then decreasing $\overline{<T>}$ yields decreasing $\Gamma$ . In other words, far from the cross-over point (from the

400    R3 to the R2 class) and in the slow rupture class, the rate of energy dissipation is low. The latter conclusion

401    has been proved numerically in [45] and recently in [46] through spring-block models. In Fig.7, we have

402    shown how the temperature of the process zone in sub-micron length scale can rise to above melting tempera-

403    ture and glass-point of $SiO_2$. The employed fracture energy remarkably does agree with the values reported in

404    [47]. **To confirm the nature of amplified events and their relations with our formulation, we tested**

405    **some tiny events from two well-studied experiments. The first experiment was a double friction test**

406    **including gouge materials sheared between two steel plates as has been described in [48]. Events emit-**

407    **ted from angular quartz sand do reflect a similar fast-slip phase duration from Lab.Eq.1 and Lab.Eq.2**

408    **(Fig.8a). The second additional experiments were from concrete samples (mainly Portland cement) as**

409    **described in [49]. They reflect about ~55-60µs duration of the fast-slip period, 2 to 3 times longer than**

410    **other silicate-based rock tests (Fig.8b). Considering the fact that the toughness of concrete is higher**

411    **than silicate-based rocks ($SiO_2$ is dominant mineral in Granite, Basalt, and some Sandstones), we assign**

412    **this feature to longer fast-slip stage. This can also be due to additional sources of energy propagation**

413    **involved in colloidal systems (such as cement and some polymers) as they include bond breaking and**

414    **bond rearrangements at the crack tip [50]. Then, our effective temperature model successfully does**

415    **match with the released energy during micro-cracking, well-coupled with the details of cracking bonds**

416    **and other additional dissipation of energy sources at the molecular/atomic levels.**

417

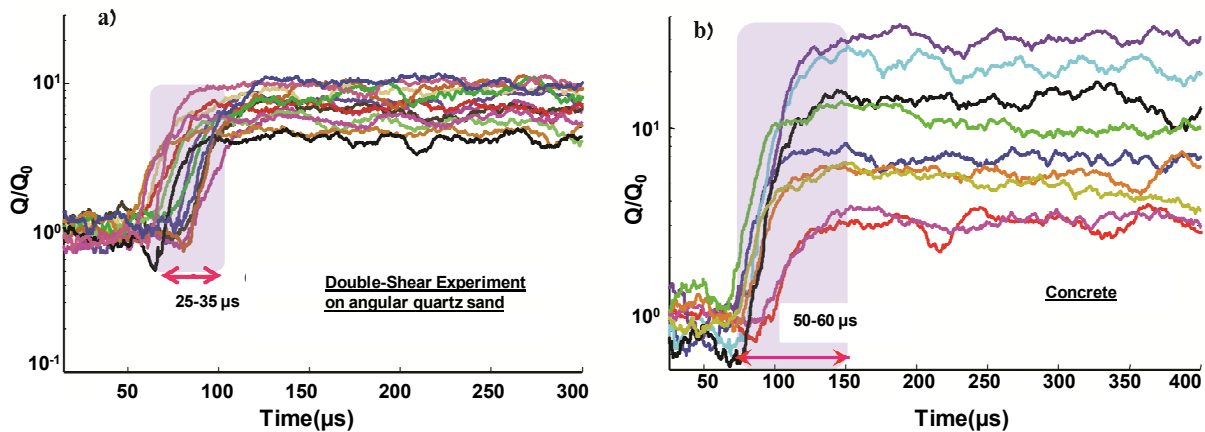

419 **Figure 8**| Duration of the second phase in two different experiments as follows: **a)** Double shear test as reported in [48] while Quartz-
420 gouge materials are sheared between two steel plates. The duration of the second phase is about 25-30µs.  **b)** Waveforms from con-
421 crete samples as reported in [49] show longer fast-slip phase. This is due to the nature of C-S-H bands, and the micro-cracks involve
422 bond breaking and bond rearrangements at the crack tip [50].

423
424
425

426    **6) Mean-field Modeling of Friction-networks -** We consider a mean-field modeling approach
427    and use the power-law nature of the load distribution (or centrality-Fig.9). To this aim, we assume a nearly
428    constant exponent of the centrality distribution: $P(BC) \sim BC^{-\Omega}, \Omega = 3$ (Fig.9a). To measure the power coeffi-
429    cient of the centrality distribution, we used likelihood estimators for fitting the power-law distribution to data
430    (Fig.9b). **Also, goodness-of-fit test using the Kolmogorov-Smirnov method was used, calculating p-value**
431    **for fitted power-law model [53]. We consider that the power law is rejected if p ≤ 0.05. For the range of**
432    $2 \leq B.C \leq 40$, **we obtained on average the aforementioned power-law distribution with p >0.3. From**
433    **Fig.9b, we learned that increasing the intensity of rupture (and then increasing the energy flow into the**
434    **interface) is roughly scaled with increasing the coefficient of the power law. Therefore, the trend of the**
435    **distribution is reciprocal to b-values as the power exponent of amplitudes distribution. Increasing b-**
436    **value indicates finding events with high amplitude (and energy) diminishes (and vice versa).**

437    Here, we build our model based on $P(BC) \sim BC^{-\Omega}$ **while we do not restrict our model to the**
438    **aforementioned interval. As the simplest algorithm and with the aid of the preferential attachment**
439    **models (such as B-A model) [51-52] to get this relation, the distribution can be interpreted as**
440    $\dfrac{\partial B.C_i}{\partial t} \propto \dfrac{B.C_i}{\sum\limits_j B.C_j}$ . **With this interpretation, we assumed network evolution is such that the nodes with**

441    **highest centrality (i.e., load) grow preferentially in terms of holding more "loads" at each step. It im-**
442    **plies that vertices with higher degrees should be loaded by allowing passage of stress-waveforms (as in-**
443    **formation) along the shortest paths [33,54]. This is an indirect inference of the evolution of networks,**
444    **while we did not directly take into account the role of addition of links as well as minimal models of**
445    **preferential attachment. Also, we note that numerical modeling proves that** *B.C* **distribution of scale**
446    **free networks follows a power law with the exponent around 2.2 which is insensitive to the exponent of**
447    **degree distribution in the range** $2 < \delta \leq 3$ ( $\delta$ **is the exponent in** $P(k) \propto k^{-\delta}$ ) **[54]. Furthermore, a re-**
448    **cent study employed the evolution of networks in terms of nodes attachment proportional to the cluster-**
449    **ing coefficients (normalized triangles) of existing node, which resulted in community formation [55]. In**
450    **the following, we find possible scenarios of edges growth in this model by using two additional relations.**
451    **We now use some of the introduced scaling relations to simplify the relation. The first scaling law re-**
452    **lates spatial mean centrality to the clustering coefficient (Fig.9c:** $< B.C >_x \propto < C >_x^{-\zeta}$ **). The exact form of the**
453    **latter relation is a power-law with cut-off parameter; however, we will not consider this fact in our cal-**
454    **culation.** The second rule connects clustering coefficient to node degree where $T(k) \propto k^{\beta}$ [28-29] ( *T* is the
455    number of triangles and *k* is the node's degree). To get the latter relation, we use $T(k) = \binom{k}{2} C_k$ and $\binom{k}{2} \sim \dfrac{k^2}{2}$ ,

456 then we find $T_T(k) \propto k^\beta$ and it eventually results in: $B.C \propto 2k^{\xi(2-\beta)}$. **Also, considering the first relation , we**

457 **find out** $<C>_s \propto <k>_s^{\beta-2}$ **where for** $\beta = 1$ **satisfies the scale free -hierarchical networks [33,61].** Then, the

458 preferential attachment is simplified as :

$$\frac{\partial k_i}{\partial t} \propto \frac{2^\xi}{A} \frac{k_i}{\sum_j k_j^A} , \qquad (9)$$

460 in which $A = \xi(2 - \beta)$ . Assuming A=1 leads to classic scale-free networks.

461 Continuing [56] to approximate a nonlinear growth of the kernel (*i.e.,* $\sum_j k_j^A$), Eq. 9 reads as an exponential

462 approximation of the links' evolution: $k_i(t) \sim \exp(\frac{-2^\xi}{A(A+1)} t^{-A+1})$. Let us assume the distribution of edges follows

463 a power law as well as $P(k) \propto k^{-\delta}$ in which $\delta = 1 - \xi\beta + 2\xi^2(\Omega-2)(2-\beta)$ is obtained by using equalizing ex-

464 pectation values of centrality and node's degree. For $\delta > 0$ (as a classic scale-free networks) and $\xi\beta > 1$ , we

465 find $\beta < 2$ . Below, we will see this finding is similar with numerical implementation of Eq.9. This numerical

466 proof rationalizes the interpretation of centrality distribution.

467

468

469

470

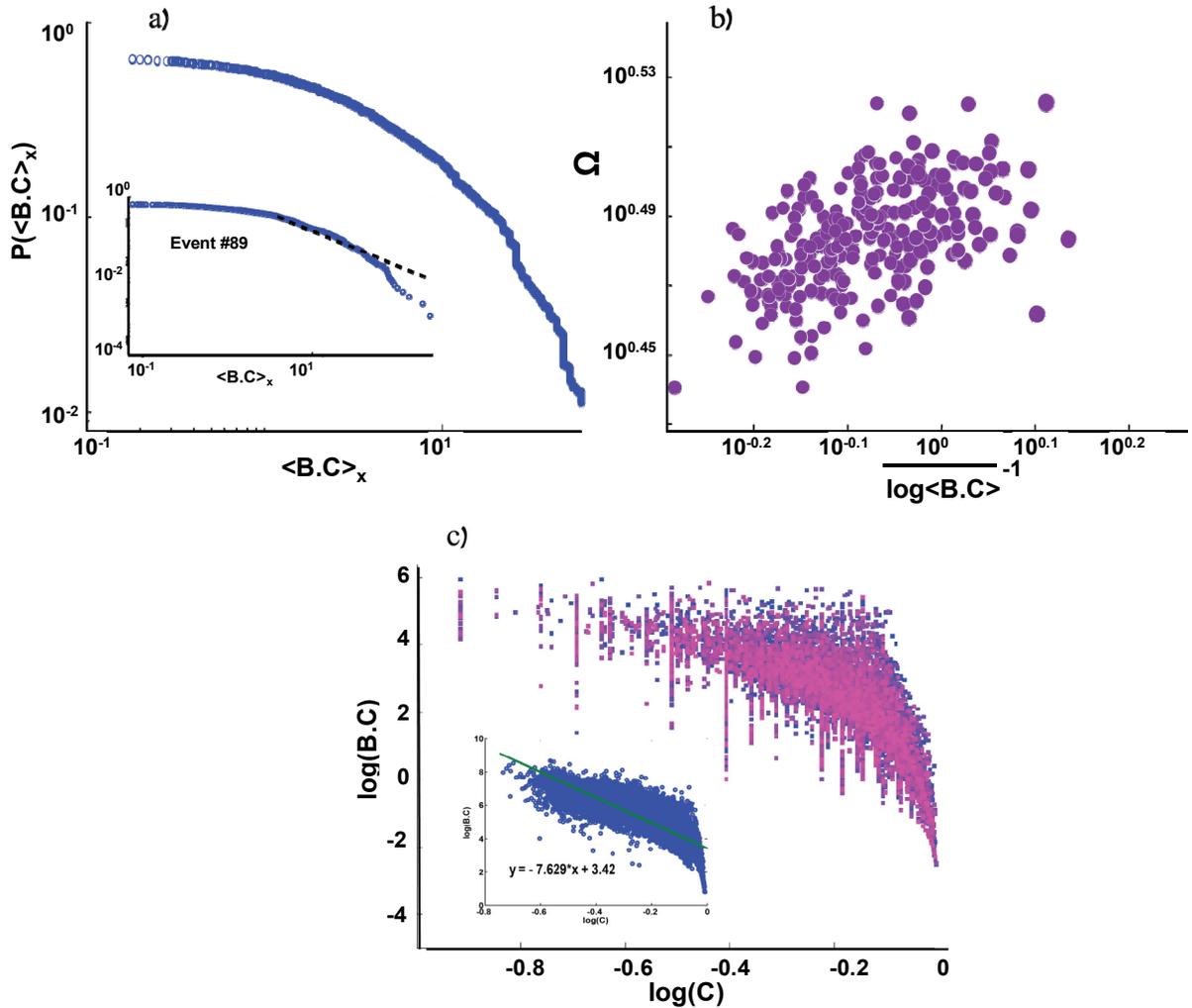

**Figure 9| Distribution of centrality. a)** Distribution of $_{<B.C>_x}$ during the duration of a precursor event # 27 from smooth fault ($t_{max.}$=204 $\mu s$) –inset shows the distribution with the best fitted line for event #89 from the Lab.Eq1. We have used algorithm described in [53] to get the power-law .**b)** The spatio-temporal average of a few hundreds of Lab.2 versus the power law coefficient in $P(BC) \sim BC^{-\Omega}$ shows that rupture fronts with higher $_{1/\overline{\log <B.C>}}$ holds higher power law coefficient. **c)** The distribution of the sampled points from an event in a log-log scale of $_{<B.C>_x - <C>_x}$. Inset shows event #25 from saw-cut experiment with the best fitted line (using Levenberg–Marquardt algorithm [60]).

We numerically simulate Eq.9, while for each time step we introduce 2 links (*m=2*). After a sufficiently long period of evolution, we evaluate the maximum node's degree in the networks while the exponent of the kernel changes (*i.e., A*)-Fig.10a. A transition is observed with increasing *A* where a systematic transition to "hubness" characteristic ($A{\rightarrow}1$) is imprinted in the positive values of the exponent (i.e., $\beta < 2$). **An expected question is whether the transition is continuous or not as *A* crosses a critical value. Our observations indicate that the transition is continuous. For example, Fig. 10a (inset) shows that as *A* ap-** 485  **proaches from above the value of** $A_{critical} \approx -5$**, the mean slope of the graph tends continuously to zero.**

486  **From this figure, we can also see that with increasing the systems size, the transition occurs in smaller**

487  $A_{critcal}$**.** This transition is equal to considering fast decay of the load distribution versus the clustering coeffi-

488  cient in which we generally expect to evaluate the state of the interface at the first phase (deformation and

489  crushing asperities-Fig.1a). With the increase of the kernel's exponent, the rich nodes have the chance to

490  grow, and the poor nodes (or weak nodes) naturally and gradually become suppressed. This is the situation of

491  "the fit-get-rich" (FGR [57])-Fig.10a.

492

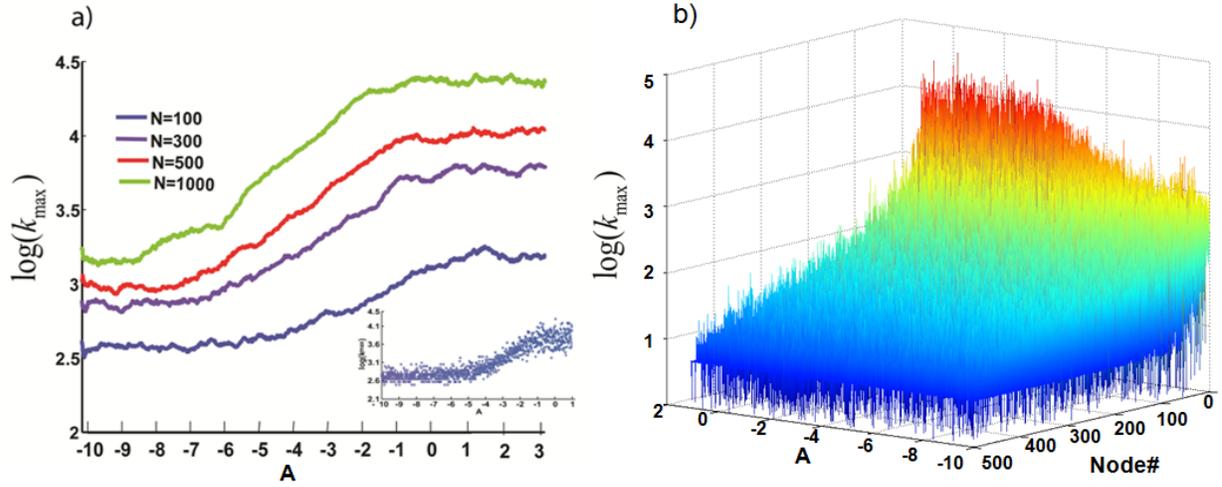

493

494  **Figure 10|**(a) The numerical simulation of Eq.9 (see the main text) shows a phase transition to FGR phase for t=100,300,500 and

495  1000. Inset shows accumulation of 10 times run of the model for N=300. (b) The distribution of the node's degree (t=500) and evolu-

496  tion of networks to the FGR phase. Older nodes gain more links as the network evolves.

497

498  Considering the similarity of Eq.(9) to the fitness model (and non-linear network models [56]), one can

499  examine the possible condensation of the phase transition (or "gel-like" transition) in the present model. One

500  of the possible scenarios is to consider a coupling term in the governing equation with the node's degree evo-

501  lution. **Assuming a non-linear network model [56], we show that (1) the node's degree is proportional to**

502  **the** $\xi$ **and (2)** $A \to 2$ **, we reach a gel-like state. The proof is as follows: to have a gel-like state, we should**

503  **satisfy:** $\dfrac{\partial k_i}{\partial t} \equiv \dfrac{k_i^A}{\sum k_j^A}, A \geq 2$**. Equalizing this condition with the Eq.9 and assuming** $A = 2$**, we reach**

504  $k_i \propto \exp(\xi - 1)$**, which can be rewritten as** $k_i \propto \exp(\dfrac{2}{2 - \beta})$**.** In other words, if the exponent of clustering-

505  centrality plane grows logarithmically proportional with the node's degree around $A=2$, we see a gel-like or

506  super-rupture state. In particular, this situation can be probed in the tail of the distribution in which we get

507 $<\xi> \propto \log A + (A-1)\log<k>$ for $A > 1$ where we considered an average value for node's degree and the expo-

508 nent. Indeed the FGR state corresponds to the R2 or R1 class. Also, we note the rapid decrease of $\beta$ (or $\chi$)

509 accelerates the growth of $k_{max}$, coinciding with the approach to the tail of the parameter space in Fig.9a. To

510 present a possible mechanical-interpretation of the condensed state of friction networks, we notice that reach-

511 ing a possible condensed state is equal to approaching a less-collective motion of atoms (i.e., rupture). In par-

512 ticular, the interface does not show collective local stick-lip motion, and the state of the motion of atoms (i.e.,

513 particles or nodes) is controlled by a "single" atom or very condensed modules of atoms. This phenomenon is

514 the condition of incommensurate contacts; they can prevent collective atomic motion (or collective particle

515 motion), leading to a "superlubricity" (or super-rupture) state of friction [58-59].

516

517 **7) Conclusions.** Novel results regarding a recent proposed approach on functional friction networks have

518 been presented. The results suggest the micron to sub-micron rupture fronts imprint critical scaling exponents

519 in the proper scalar network's parameter spaces, comparable with the possible slow and regular deformation

520 or "earthquakes". Further analysis of the renormalization group will complete the presented parameter spaces.

521 To further the interpretation of the functional networks, two different scenarios were introduced. The first ap-

522 proach featured analysis of the energy of the networks based on the change of modularity index. This enabled

523 inference of a possible relationship between the remote stress field as the driving field and the temporal aver-

524 age modularity comparable with the quantified experimental results.

525 A singularity in the transition from slow ruptures to sub-Rayleigh class (or vice versa) has been demonstrated,

526 compatible with the recent direct observation of rupture cross-over in glassy materials. A second model used a

527 mean-filed approximation of networks' evolution with well-established scale-free networks to simulate the

528 evolution of a node's degree. The governing equation was the result of the collective distribution of events in

529 different network parameter spaces. The model described a transition period to scale free or FGR phases, de-

530 pending on the values of the two variables of the model. We also studied the gel-like state of the functional

531 friction networks and speculate that this special case can be compared with super-ruptures or a super-lubricity

532 state of friction. Obviously, the latter finding needs more investigation regarding recorded waveforms in su-

533 per-shear or super-Rayleigh cases. Collectively, this work demonstrates network theory has a great potential

534 to analyze and extract new information from laboratory experiments featuring full waveform AEs. This data

535 can be used to model the mechanical phenomena on macroscopic and sub-micron (and below) scales. The in-

536 terpretation of network parameters in the light of nearly constant evolutionary phases has the potential to be a

537 very useful indicator and provide new insights into the evolution of rupture tips. The techniques and ap-

538 proaches herein described have great potential in terms of their application to AE events originating from

539 conditions such as complex boundary conditions, fluid-driven fractures, noise coupled phenomena such as 540 thermal–mechanical cracks, events from high temperature-high pressure (multi-anvil tests) and crackling nois-
541 es from sheared granular materials. Our results also hint at drawing universal mechanisms of acoustic emis-
542 sions in terms of their main evolutionary phases. While materials mainly involved $SiO_2$ minerals reflect nearly
543 an invariant fast-slip phase in their crackling noises under normal loading conditions, materials with more
544 complex molecular structures (such as cement, bone, protein materials) mark longer rising times. This indicat-
545 ed that acoustic waveforms can shed light on some features of complex failure of atomic bonds.  Furthermore,
546 from network science perspective, this research featured relatively simple algorithms in the construction of the
547 featured networks. Future approaches include the use of more sophisticated, weighted networks and directed
548 graph theory to better elucidate the behavioral processes and transitions from micro to macro-scale fractures.
549
550
551 **Acknowledgements**
552

553     We would like to acknowledge and thank D. Lockner (USGS, Menlo-Park, USA) , Stefan Nielsen (Istituto Nazionale di Geofisica e Vulcanologia,
554 Italy) , Tatyana Katsaga (Itasca, Toronto) , Karen Mair (Physics of Geological Processes, University of Oslo, Oslo, Norway ) and C.Marone
555 (Pennsylvania State University, University Park, Pennsylvania, USA) for providing  the data set employed in this work.  The first author would like to
556 acknowledge Prof. K.Xia 's comments and encouragements during the preparation of the manuscript.   Reviewers' comments helped to increase the
557 quality of the initial manuscript and are gratefully acknowledged. The first author would like to acknowledge comments and points from Dr. Reik
558 Donner.

559 **Author Contributions**
560  H.O.G.H. performed the calculations and wrote the manuscript. B.D. accomplished and designed the tests and R.P.Y. supervised the research and pro-
561 vided other employed data sets . All authors discussed and commented on the results.
562
563 **References**

564  [1] Meng, L. Inbal, A. and Ampuero, J.-P.: A window into the complexity of the dynamic rupture of the 2011 Mw 9 Tohoku-Oki earthquake, **Geophys.**
565 **Res. Lett,** 38, L00G07, 2011.
566 [2] Maercklin, N. Festa, G. Colombelli, S. and  Zollo, A. :Twin ruptures grew to build up the giant 2011 Tohoku, Japan, earthquake, **Scientific reports**,
567 2, 2012.
568 [3] Simons, M. Minson, S. E. Sladen, A. Ortega, F. Jiang, J. Owen, S. E. and Webb, F. H.: The 2011 magnitude 9.0 Tohoku-Oki earthquake: Mosaicking
569 the megathrust from seconds to centuries, **Science**, 332(6036), 1421-1425, 2011.
570 [4] Rubinstein ,S., Cohen, G. and Fineberg, J.: Detachment fronts and the onset of dynamic friction, **Nature.** 430, 1005-1009 ,2004.
571 [5] Ben-David, O. Cohen,G. & Fineberg,J.: The dynamics of the onset of frictional slip. **Science**. 330, 211 ,2010.
572 [6] Ben-David, O. Rubinstein,S. &  Fineberg ,J.: Slip-Stick: The evolution of frictional strength. **Nature** .463, 76 ,2010.
573 [7] Ghaffari, H. O. & Young, R. P. : Acoustic-friction networks and the evolution of precursor rupture fronts in laboratory earthquakes, **Scientific**
574 **Reports** 3, 1799, doi:10.1038/srep01799,2013.
575 [8] Nielsen ,S., Taddeucci, J. & Vinciguerra, S.: Experimental observation of stick-slip instability fronts. **Geophys. J. Int.** 180, 697 ,2010.
576  [9] Andrews, D. J. : Rupture velocity of plane strain shear cracks. **J. Geophys. Res. B** 81, 5679–5687 ,1976.
577 [10] Dunham, E.M.: Conditions governing the occurrence of supershear ruptures under slip-weakening friction. **J. Geophys. Res.** 112, B07302 ,2007.
578 [11] Rice, J. R., Lapusta, N. & Ranjith, K.: Rate and state dependent friction and the stability of sliding between elastically deformable solids. **J. Mech.**
579     **Phys. Solids** 49, 1865–1898 ,2001.
580 [12] Noda, H., & Lapusta, N.:  Stable creeping fault segments can become destructive as a result of dynamic weakening. **Nature,** 493(7433), 518-
581 521,2013.

[13] Segall, P., Rubin, A. M., Bradley, A. M. & Rice, J. R. : Dilatant strengthening as a mechanism for slow slip events. **J. Geophys. Res.** 115, B12305 ,2010.

[14] Kaneko, Y., & Ampuero, J. P.: A mechanism for preseismic steady rupture fronts observed in laboratory experiments. **Geophysical Research Letters**, 38(21), 2011.

[15] Ben-David, O., G. Cohen, and J. Fineberg.: Short-time dynamics of frictional strength in dry friction, **Tribology Letters** 39.3 235-245,2010.

[16] Thompson,B.D., Young, R.P.& Lockner, D.A.: Observations of premonitory acoustic emission and slip nucleation during a stick slip experiment in smooth faulted Westerly granite. **Geophys. Res. Lett.** 32 L10304 ,2005.

[17] Thompson, B.D., Young, R.P.& Lockner, D.A.: Premonitory acoustic emissions and stick-slip in natural and smooth-faulted Westerly granite. **J Geophys Res**. 114, B02205J ,2009.

[18] Donner, R. V., Zou, Y., Donges, J. F., Marwan, N., & Kurths, J. : Recurrence networks—A novel paradigm for nonlinear time series analysis. **New Journal of Physics**, 12(3), 033025, 2010.

[19] Bassett, D. S., Meyer-Lindenberg, A., Achard, S., Duke, T., & Bullmore, E.: Adaptive reconfiguration of fractal small-world human brain functional networks. **PNAS**, 103(51), 19518-19523, 2006.

[20] Donges, J. F., Zou, Y., Marwan, N., & Kurths, J.: The backbone of the climate network. **EPL (Europhysics Letters)**, 87(4), 48007, 2009.

[21] Li, C-B, Yang, H., and Komatsuzaki,T. :Multiscale complex network of protein conformational fluctuations in single-molecule time series." **Proceedings of the National Academy of Sciences** 105.2 536-541,2008.

[22] Bashan, A., Bartsch, R. P., Kantelhardt, J. W., Havlin, S., & Ivanov, P. C. Network physiology reveals relations between network topology and physiological function. **Nature communications**, 3, 702, 2012.

[23] Schiff, S. J., So, P., Chang, T., Burke, R. E. & Sauer, T. : Detecting dynamical interdependence and generalized synchrony through mutual prediction in a neural ensemble. **Phys. Rev. E** 54, 6708–6724 ,1996.

[24] Arnhold, J., Grassberger, P., Lehnertz, K. & Elger, C. E. : A robust method for detecting interdependences: application to intracranially recorded EEG. **Physica D** 134, 419–430 ,1999.

[25] Chen, Y., Rangarajan, G., Feng, J. & Ding, M.: Analyzing multiple nonlinear time series with extended Granger causality. **Phys. Lett. A** 324, 26–35 ,2004.

[26] Napoletani, D. & Sauer, T.: Reconstructing the topology of sparsely connected dynamical networks. **Phys. Rev. E** 77, 026103 ,2008

[27] Iwayama, K., Hirata, Y., Takahashi, K., Watanabe, K., Aihara, K.& Suzuki, H.: Characterizing global evolutions of complex systems via intermediate network representations, **Scientific Reports** 2, 423, 2012.

[28] Ghaffari , H. O. & Young, R. P. :Network configurations of dynamic friction patterns, **EPL** 98 (4), 48003 ,2012.

[29] Ghaffari, H. O. & Young, R. P. :Topological complexity of frictional interfaces: friction networks, **Nonlinear Processes Geophys**.19, 215,2012.

[30] Newman, M. E. J. & Girvan, M.: Finding and evaluating community structure in networks. **Phys. Rev. E**, 69, no. 026113 ,2004.

[31] Fortunato, S. : Community detection in graphs. **Phys. Rep.** 486, 75–174 ,2010.

[32] Guimerà, R., & Amaral, L. A. N.: Functional cartography of complex metabolic networks. **Nature.** 433, 895–900 ,2005.

[33] Newman, M.E.J.: Networks: An Introduction ( Oxford University Press, 2010).

[34] Mehta, A. P., Mills, A. C., Dahmen, K. A., & Sethna, J. P. : Universal pulse shape scaling function and exponents: Critical test for avalanche models applied to Barkhausen noise. **Physical Review E**. 65(4), 046139 ,2002.

[35] Zapperi, S., Castellano, C., Colaiori, F. & Durin, G. :Signature of effective mass in crackling-noise asymmetry. **Nature Phys.** 1, 46–49 ,2005.

[36] Sone, H., & Shimamoto, T.: Frictional resistance of faults during accelerating and decelerating earthquake slip, **Nature Geoscience**, 2(10), 705-708, 2009.

[37] Andrews, D. J. : Dynamic plane-strain shear rupture with a slip-weakening friction law calculated by a boundary integral method, **Bulletin of the Seismological Society of America** 75.1 1-21,1985.

[38] Palla, G., Derényi, I., Farkas, I., & Vicsek, T.: Statistical mechanics of topological phase transitions in networks. **Physical Review E**, 69(4), 046117,2004.

[39] Rycroft, C. H., & Bouchbinder, E. : Fracture Toughness of Metallic Glasses: Annealing-Induced Embrittlement. **Physical Review Letters**, 109(19), 194301,2012.

[40] Xu, X., Zhang, J., & Small, M.: Superfamily phenomena and motifs of networks induced from time series. **PNAS**. 105(50), 19601-19605,2008.

[41] Milo, R., Shen-Orr, S., Itzkovitz, S., Kashtan, N., Chklovskii, D., & Alon, U. : Network motifs: simple building blocks of complex networks. **Science**, 298(5594), 824-827,2002.

[42] Alon, U.: Network motifs: theory and experimental approaches. **Nature Reviews Genetics**, 8(6), 450-461,2007.

[43] Freund, L. B. :The mechanics of dynamic shear crack propagation. **J. Geophys. Res.** 84, 2199–2209 ,1979.

[44] Livne, A., Bouchbinder, E., and Fineberg, J.: The breakdown of linear elastic fracture mechanics near the tip of a rapid crack. **Phys. Rev. Lett.** 101, 264301 ,2008.

[45] Braun, O. M., Barel, I. & Urbakh, M. :Dynamics of transition from static to kinetic friction. **Phys. Rev. Lett,.** 103, 2009.

[46] Bouchbinder, E. Brener, E. A., Barel, I.& Urbakh, M.: Dynamics at the onset of frictional sliding. **Phys. Rev. Lett**. 107, 235501 ,2011.

[47] Lockner, D. A., Byerlee, J. D., Kuksenko, V., Ponomarev, A., & Sidorin, A.: Quasi-static fault growth and shear fracture energy in granite. **Nature**,350(6313), 39-42,1991.

[48] Mair, K., Marone, C., & Young, R. P.: Rate dependence of acoustic emissions generated during shear of simulated fault gouge, **Bulletin of the Seismological Society of America**, 97(6), 1841-1849, 2007.

[49] Katsaga, T., Sherwood, E. G., Collins, M. P., & Young, R. P. : Acoustic emission imaging of shear failure in large reinforced concrete structures, **International Journal of Fracture**, 148(1), 29-45,2007.

[50] Buehler, M. J., & Keten, S.: Colloquium: Failure of molecules, bones, and the Earth itself. **Reviews of Modern Physics**, 82(2), 1459, 2010.

[51] Albert, R., & Barabási, A. L.: Statistical mechanics of complex networks. **Reviews of modern physics**, 74(1), 47,2002.

[52] Barabási, A. L., Albert, R., & Jeong, H.: Mean-field theory for scale-free random networks. **Physica A: Statistical Mechanics and its Applications**, 272(1), 173-187,1999.

[53] Clauset, A., Shalizi, C.R. and Newman, M.E.J.: Power-law distributions in empirical data, **SIAM Review**, 51(4), 661-703 ,2009.

[54] Goh, K-I., B. Kahng, and D. Kim. :Universal behavior of load distribution in scale-free networks. **Physical Review Letters** 87.27 ,2001.

[55] Bagrow, J-P., and Brockmann ,D.: Natural emergence of clusters and bursts in network evolution, **Physical Review X**, 3.2, 021016,2013.

[56] Krapivsky, P. L.; Redner, S.; Leyvraz, F. (2000). :Connectivity of Growing Random Networks. **Phys. Rev. Lett**. 85: 4629–32.

[57] Bianconi, G.; Barabási, A.-L. : Bose–Einstein Condensation in Complex Networks, **Phys. Rev. Lett**. 86: 5632–35, 2001.

[58] Shinjo, K., and Hirano, M.: Dynamics of friction: superlubric state. **Surface Science**, 283(1), 473-478,1993.

[59] Dienwiebel, M., Verhoeven, G. S., Pradeep, N., Frenken, J. W., Heimberg, J. A., and Zandbergen, H. W. : Superlubricity of graphite, **Physical review letters**, 92(12), 126101, 2004.

[60] Levenberg, K.: A Method for the Solution of Certain Non-Linear Problems in Least Squares, **Quarterly of Applied Mathematics** 2: 164–168,1944.

[61] Ravasz, Erzsébet, et al.: Hierarchical organization of modularity in metabolic networks., **Science** 297.5586 1551-1555, 2002.